\documentclass[aps,pre,twocolumn,superscriptaddress,showpacs,showkeys ]{revtex4-1}

\usepackage{graphicx}                   
\usepackage{hyperref}                   
\usepackage{amsmath,amssymb}            
\usepackage{epstopdf}
\usepackage{subcaption}					

\usepackage{color}
\definecolor{orange}{rgb}{1,0.5,0}
\definecolor{brown}{rgb}{0.65, 0.16, 0.16}
\definecolor{phlox}{rgb}{0.87, 0.0, 1.0}

\begin{document}

\title{Dynamical Crossover in Invasion Percolation}

\author{S. Tizdast}
\affiliation{Department of Physics, University of Mohaghegh Ardabili, P.O. Box 179, Ardabil, Iran}

\author{N. Ahadpour}
\affiliation{Department of Physics, University of Mohaghegh Ardabili, P.O. Box 179, Ardabil, Iran}

\author{M. N. Najafi}
\affiliation{Department of Physics, University of Mohaghegh Ardabili, P.O. Box 179, Ardabil, Iran}
\email{morteza.nattagh@gmail.com}

\author{Z. Ebadi}
\affiliation{Department of Physics, University of Mohaghegh Ardabili, P.O. Box 179, Ardabil, Iran}

\author{H. Mohamadzadeh}
\affiliation{Department of Physics, University of Mohaghegh Ardabili, P.O. Box 179, Ardabil, Iran}

\begin{abstract}
The dynamical properties of the invasion percolation on the square lattice are investigated with emphasis on the geometrical properties on the growing cluster of infected sites. The exterior frontier of this cluster forms a critical loop ensemble (CLE), whose length $(l)$, the radius $(r)$ and also roughness $(w)$ fulfill the finite size scaling hypothesis. The dynamical fractal dimension of the CLE defined as the exponent of the scaling relation between $l$ and $r$ is estimated to be $D_f=1.81\pm0.02$. By studying the autocorrelation functions of these quantities we show importantly that there is a crossover between two time regimes, in which these functions change behavior from power-law at the small times, to exponential decay at long times. In the vicinity of this crossover time, these functions are estimated by log-normal functions. We also show that the increments of the considered statistical quantities, which are related to the random forces governing the dynamics of the observables undergo an anticorrelation/correlation transition at the time that the crossover takes place.
\end{abstract}


\maketitle

\section{Introduction}
Multiphase flow phenomena in porous media is of great importance in many fields of science and industry. A very promising and highly-used strategy for analyzing this is the cavity lattice method, used commonly in statistical models, expressing the porous media parameters as the local fields, which enables one to use the concepts of the percolation theory in fluid flow modeling. Among the models that aim to describe two-phase flow in porous media, invasion percolation (IP) has been subjected to intense studies due to its relative simple structure (making it suitable for statistical investigations) and its success in capturing most relevant physical processes, and also showing a plenty interesting behaviors, like self-organizing in a critical state~\cite{wilkinson1984monte}. IP is a standard model to study the dynamics of two immiscible phases (commonly denoted by wet and non-wet phases) in a porous medium~\cite{glass1996simulation,sheppard1999invasion}. During this process, the wet phase invades the non-wet phase, and the front separating the two fluids advances by invading the pore throat at the front with the lowest threshold~\cite{sheppard1999invasion}. Many aspects of IP is known, involving various fractal dimentions~\cite{wilkinson1984monte,schwarzer1999structural,sheppard1999invasion,lopez2003fractal}, its properties in three dimensions~\cite{xu2008dynamics}, its fractal growth~\cite{pietronero1990invasion}, its dependence on the coordination number~\cite{knackstedt2002nonuniversality}, the effects of long-range correlations~\cite{prakash1992structural,knackstedt2000invasion}, and its dynamics in correlated porous media~\cite{vidales1996invasion,babadagli2000invasion}. It has also many applications in the reservoir engineering~\cite{peter2018percolation}, and also geoscience~\cite{hunt2017fractals}. Furthermore the ideas of IP been moved to other fields, like the frustrated models, in which the dynamics is due to invaded clusters~\cite{machta1995invaded,franzese1998invaded}. The dynamical studies on IP show also a rich structure. For example, the spatiotemporal properties of this model was studied in~\cite{furuberg1988dynamics}, where some dynamical scaling behaviors were found. Mapping to a fractal growth process is an interesting strategy, that was used in~\cite{pietronero1990invasion}, in which the obtained the mass fractal dimension $D=1.887$ whose exact value is $\frac{91}{48}=1.8958...$, for review see~\cite{isichenko1992percolation} and~\cite{saberi2015recent}. \\

Despite of its simple rules and structure, the IP model occasionally surprises us with some new novel features and properties. Here we focus on its temporal dynamical properties, and uncover a new dynamical crossover. The temporal dynamics of IP based on the burst dynamics was first proposed and analyzed in~\cite{roux1989temporal}, where using its connection to the ordinary percolation theory various exponents were derived. Ref.~\cite{furuberg1988dynamics} was the first to propose a scaling between length and time in IP by investigating the distribution function $P(r,t)$, $r$ being the distance between two sites added to the invaded cluster and separated by a time interval of $t$, and found that $P(r,t)\sim r^{-1}\phi(r^D/t)$, where $D\approx 1.82$ refers to the fractal dimension of the invaded cluster~\cite{furuberg1988dynamics,roux1989temporal}, as was also obtained for IPs in the low coordination number regime~\cite{knackstedt2002nonuniversality}. Here we obtain the dynamical fractal dimensions of the growing clusters. We also report on a dynamical crossover, which has not been seen in the previous studies. We show that the temporal autocorrelations of the statistical observables change behavior at a crossover time $t_{\text{crossover}}$ from power-law to exponential decay. We also show that the effective random forces that govern the growth of the invading-phase cluster also change dramatically behavior at this point. Importantly their autocorrelations change sign, showing that the system undergoes a anticorrelation/correlation crossover.\\
\\
The paper has been organized as follows: In the next section we shortly introduce the model. The section~\ref{SEC:results} has been devoted to the numerical details and results. We close the paper by a conclusion.

\section{The model}\label{Definition}
In this model, one starts by slow injection of a wetting fluid into a horizontal cell saturated with a denser nonwetting fluid,  often called the defender. The invading fluid moves preferentially through the pores with the least resistance. In the IP with (without) trapping, and the defender is supposed to be an incompressible (compressible) fluid, so that if a bubble is surrounded by an invader, it becomes impenetrable (permeable) to the invading phase~\cite{dias1986percolation}. In this work, we consider no trapping. We close the paper by a conclusion~\ref{SEC:concl}.\\

Let us describe the invasion percolation on a $L\times L$ square lattice. We assign uncorrelated uniform random numbers in the interval $ \left[ 0,1\right]$ to all lattice sites and choose the site $\textbf{r}_0=(\frac{L}{2},\frac{L}{2})$ as the seed of the growth, i.e. for injecting fluid. The random numbers represent the resistance of the pores, so that the fluid moves preferentially through the regions with the least resistance, i.e. minimum $r$. The dynamics starts with injecting fluid to $r_0$. At each time step, a unique neighbor site with smallest associated random number $r$ is occupied. In case of degeneracy, i.e. there is a set of sites with the same (lowest) $r$, we choose a site from the set randomly. As the time goes on, a connected cluster of occupied sites forms, which we call occupied sites cluster (OSC). In each sample, the dynamics is represented by a \textit{time} parameter, defined by $t\equiv\left[\frac{m}{10} \right] $, where $m$ is the mass (the number of occupied sites) of the growing cluster at that time, and $\left[ \right] $ is the integer part. $t_{\text{perc}}$ is defined as the time in which two opposite boundaries are connected by a giant cluster, when the process stops. In this time, OSC is a giant cluster which spans the lattice (not all sites), namely spanning OSC (SOSC). There is however another time scale in between, denoted by $t_{\text{BH}}$, in which the invading cluster hits one of the boundaries for the first time. Its importance is in the fact that for the times larger than it, the geometrical properties of OSC changes due to the boundary. Periodic boundary condition is considered for one direction, and open boundary condition is imposed for the boundaries in the perpendicular direction.\\

\begin{figure}
	\centerline{\includegraphics[scale=.55]{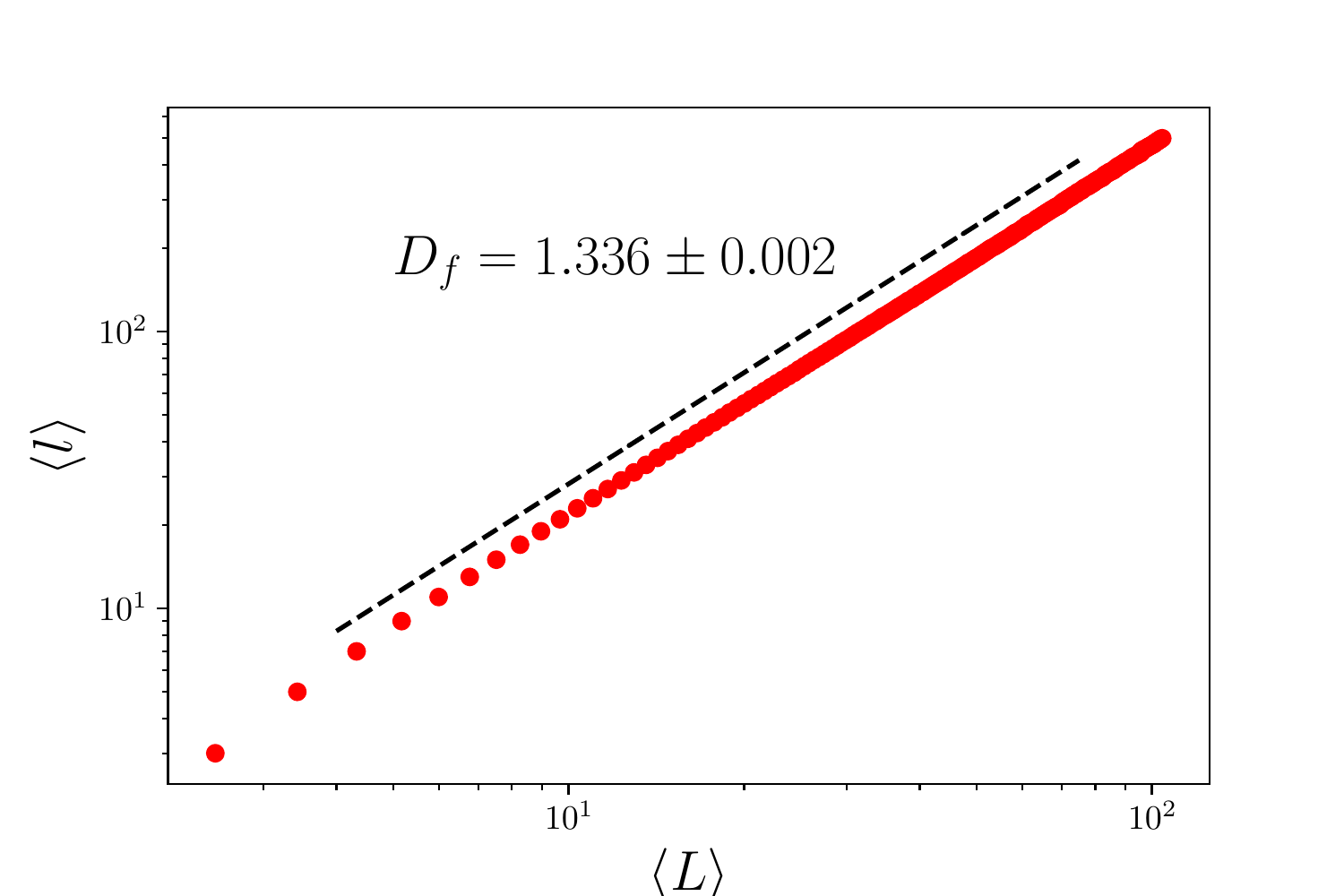}}
	\caption{Log-log plot of the $l-r_l$ graph for the largest hole using the box-counting method. (not to be confused with the dynamical fractal dimension). The slope is the fractal dimension, which is estimated by $1.33\pm 0.002$.}
	\label{fig:bd}
\end{figure}

At each time gyration radius ($r_l$), the mass gyration radius $r_m$ and the length ($l$) and the roughness ($w$) of the external perimeter of the OSC are recorded for each time. These quantities are defined as $r_l\equiv\sqrt{\frac{1}{l}\sum_{i=1}^l\left|\textbf{X}_i-\bar{\textbf{X}}\right|^2 }$, $r_m\equiv\sqrt{\frac{1}{m}\sum_{i=1}^m\left|\textbf{X}_i-\bar{\textbf{X}}\right|^2 }$  where the sums run over the sites on the boundary ($r_l$) and total sites ($r_m$) of the OSC, and $\textbf{X}_i\equiv (x_i,y_i)$, and $x_i$ and $y_i$ are the Cartesian coordinates of the site $i$, and
\begin{equation}
w^2 =  \frac{1}{l}\sum_{i = 1}^l\left(r_i-\bar{r} \right)^2
\end{equation}
where again the sum runs over the external perimeter of OSC, $r_i\equiv (x_i^2 + y_i^2)^{1/2}$ and $\bar{r}\equiv \frac{1}{l}\sum_{i = 1}^l\left[ (x_i^2 + y_i^2)^{1/2}\right]$. The the external perimeter of a OSC, may be closed loop (for finite OSCs) or be open (for SOSC or boundary-hit OSC). For the former case the fractal dimension is defined as $\left\langle \log l\right\rangle=D^I_f\left\langle\log r_l \right\rangle +cnte $ ($\left\langle \right\rangle $ being the ensemble average), whereas for the latter case we use the box-counting scheme to find $D^I_f$. The holes of the system are obtained using the Hoshen Kopelman (HK) algorithm~\cite{hoshen1976percolation}. Importantly, using HK algorithm we extracted and analyzed the \textit{largest hole} at the $t_{\text{perc}}$ for which the fractal dimension was obtained to be $D_f=1.33\pm 0.01$, as depicted in Fig.~\ref{fig:bd}. It is consistent with the fractal dimension of the self-avoiding walks $D_f^{SAW}=\frac{4}{3}$~\cite{cheraghalizadeh2018self,cheraghalizadeh2017mapping}, which is related to the fractal dimension of the interfaces of percolation $D_f^{\text{percolation}}=\frac{7}{4}$ by $\left( D^{H}_f-1\right) \left(D^I_f-1 \right)=\frac{1}{4}$~\cite{duplantier2000conformally}, where $D^I_f$ is the fractal dimension of the original traces, and $D^{H}_f$ is the fractal dimension of their hull. This is also understood in terms of Schramm-Loewner evolution (SLE) duality, stating that the diffusivity parameter $\kappa$ of a SLE class is dual to another SLE with the diffusivity parameter $\bar{\kappa}\equiv \frac{16}{\kappa}$~\cite{Najafi2012Observation,najafi2015fokker,najafi2013left}. This operation does not change the CFT universality class of the model, as a well-known result for IP~\cite{sheppard1999invasion}.
We take the ensemble averages up to $t_{\text{perc}}$, and obtain various dynamical statistical observables. Our main focus is on the spatiotemporal properties of the model, and also temporal structure of noises. To this end we concentrate of the temporal dynamics of the growing interfaces that separate wetting-nonwetting phases, as well as the bulk properties. The auto correlation functions are suitable quantities to reveal the temporal structure of the observables as time series. Consider the normalized quantities
\begin{equation}
f_x(t)=\frac{x(t)}{\left\langle x(t)\right\rangle} -1
\end{equation}
where $x=r_l,l,w$ are our statistical observables. These functions have been built in such a way that $\left\langle f_i(t)\right\rangle  =0$. Also let us consider the increments $\Delta f_x(t)\equiv f_x(t)-f_x(t-1)$, which is the noise associate with the quantity $x$. The autocorrelation is defined by 
\begin{equation}
\begin{split}
\text{Auto}_x (t,\tau)&\equiv \left\langle f_x(t)f_x(t+\tau)\right\rangle \\
\text{Auto}^{\Delta}_x (t,\tau)&\equiv \left\langle \Delta f_x(t)\Delta f_x(t+\tau)\right\rangle 
\end{split}	
\end{equation}
Note that for all the quantities that we considered, the process is not stationary so that $\text{Auto}_x (t,\tau)\neq \text{Auto}_x(t-\tau)$. Therefore, the power spectrum is more complicated than the stationary case (for which the power spectrum is simply the Fourier transform with respect to $\tau$). For later convenience, let us also define the cumulative autocorrelation function as follows:
\begin{equation}
\text{Auto}^C_x (t,\tau)= \sum_{\tau'=1}^{\tau}\text{Auto}_x (t,\tau')
\end{equation}
For the case where $\text{Auto}_x$ follows a power-law behavior with respect to $\tau$, i.e. $\text{Auto}_x (t,\tau)=c(t)\tau^{-\zeta_x}$ ($c(t)$ being an arbitrary function of $t$), then one can easily show that $\text{Auto}^C_x (t,\tau)=c(t)H(\tau,\zeta_x)$, where $H(n,r)$ is the $n^{\text{th}}$ harmonic number of order $r$. This function, helps to decide about the noisy functions by killing the noise, to firstly observe if they are fitted to power-law functions, and secondly calculate the corresponding exponent. We will observe that the cumulative autocorrelation functions are properly smooth allowing us to detect and investigate the power-law behaviors.
\begin{figure*}
	\begin{subfigure}{0.45\textwidth}\includegraphics[width=\textwidth]{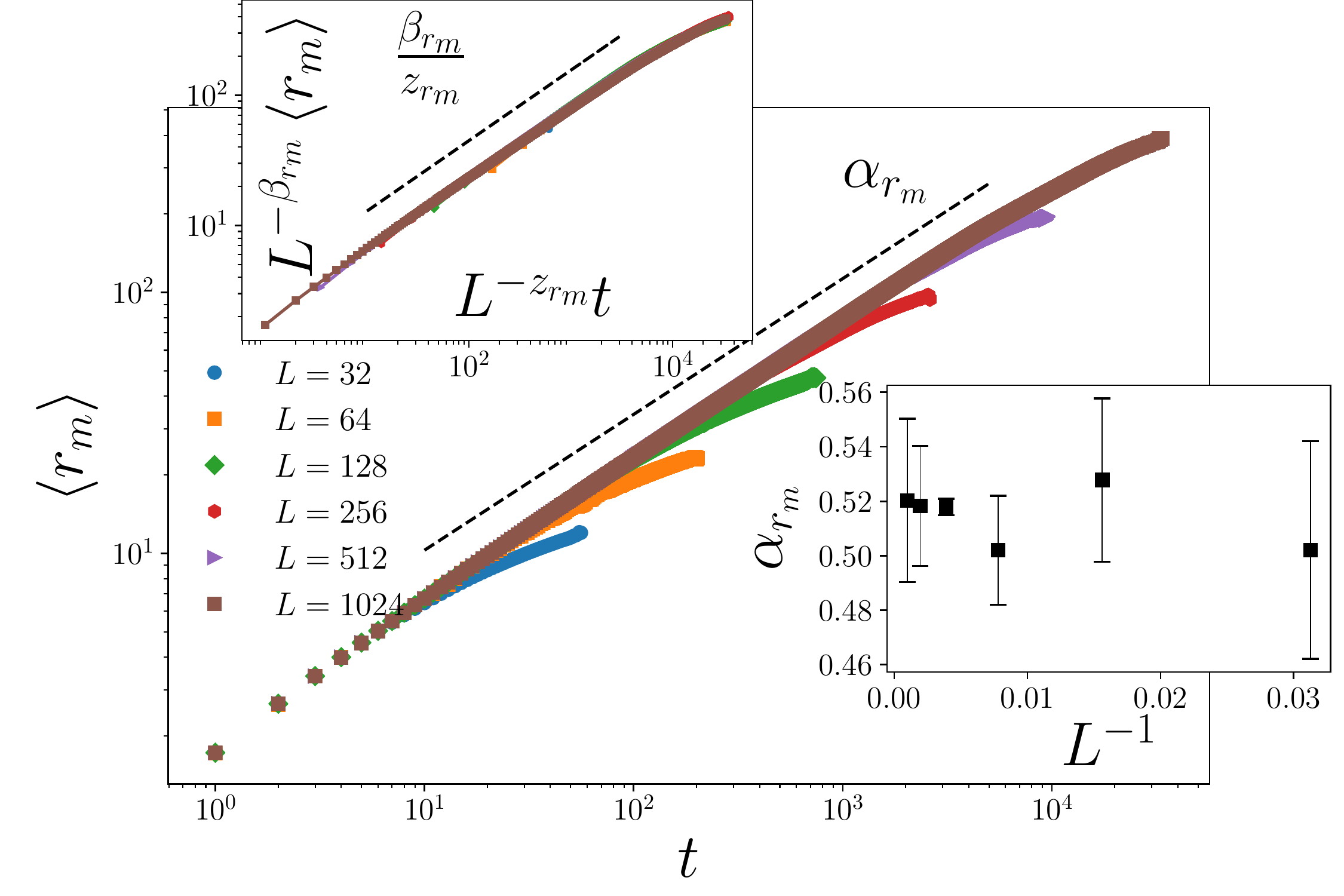}
		\caption{}
		\label{fig:t_rm}
	\end{subfigure}
	\begin{subfigure}{0.45\textwidth}\includegraphics[width=\textwidth]{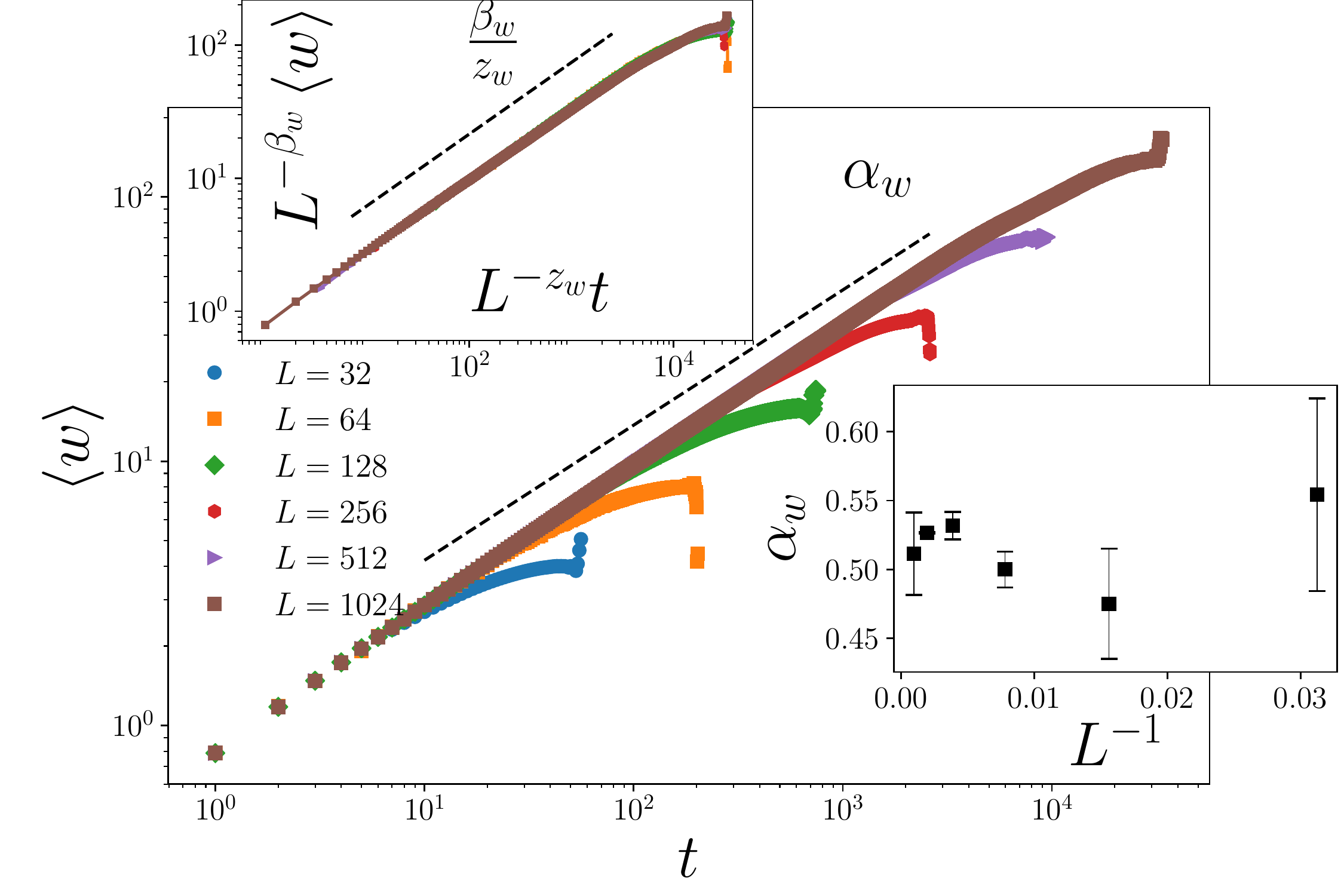}
		\caption{}
		\label{fig:t_w}
	\end{subfigure}
	\begin{subfigure}{0.45\textwidth}\includegraphics[width=\textwidth]{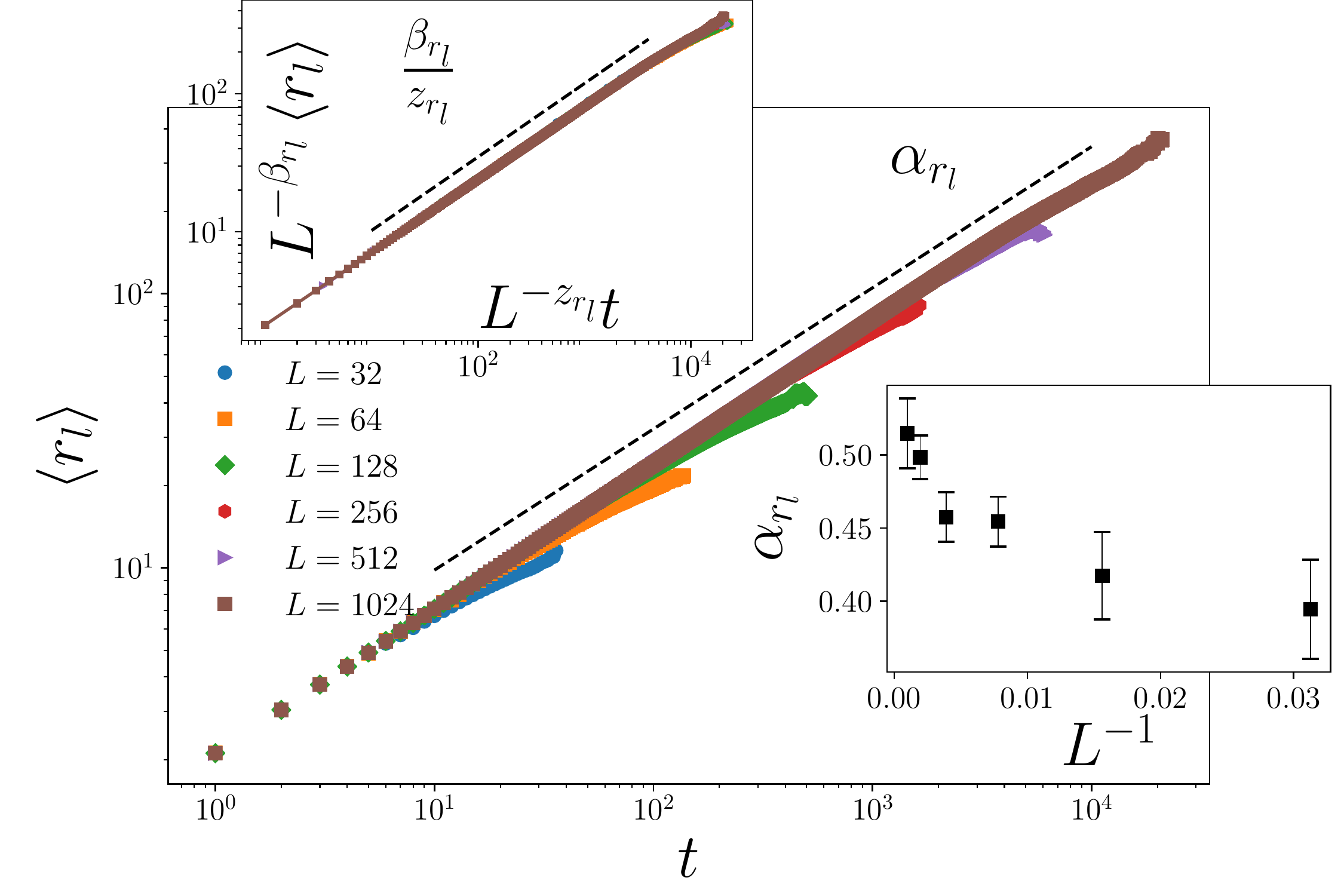}
		\caption{}
		\label{fig:t_rl}
	\end{subfigure}
	\begin{subfigure}{0.45\textwidth}\includegraphics[width=\textwidth]{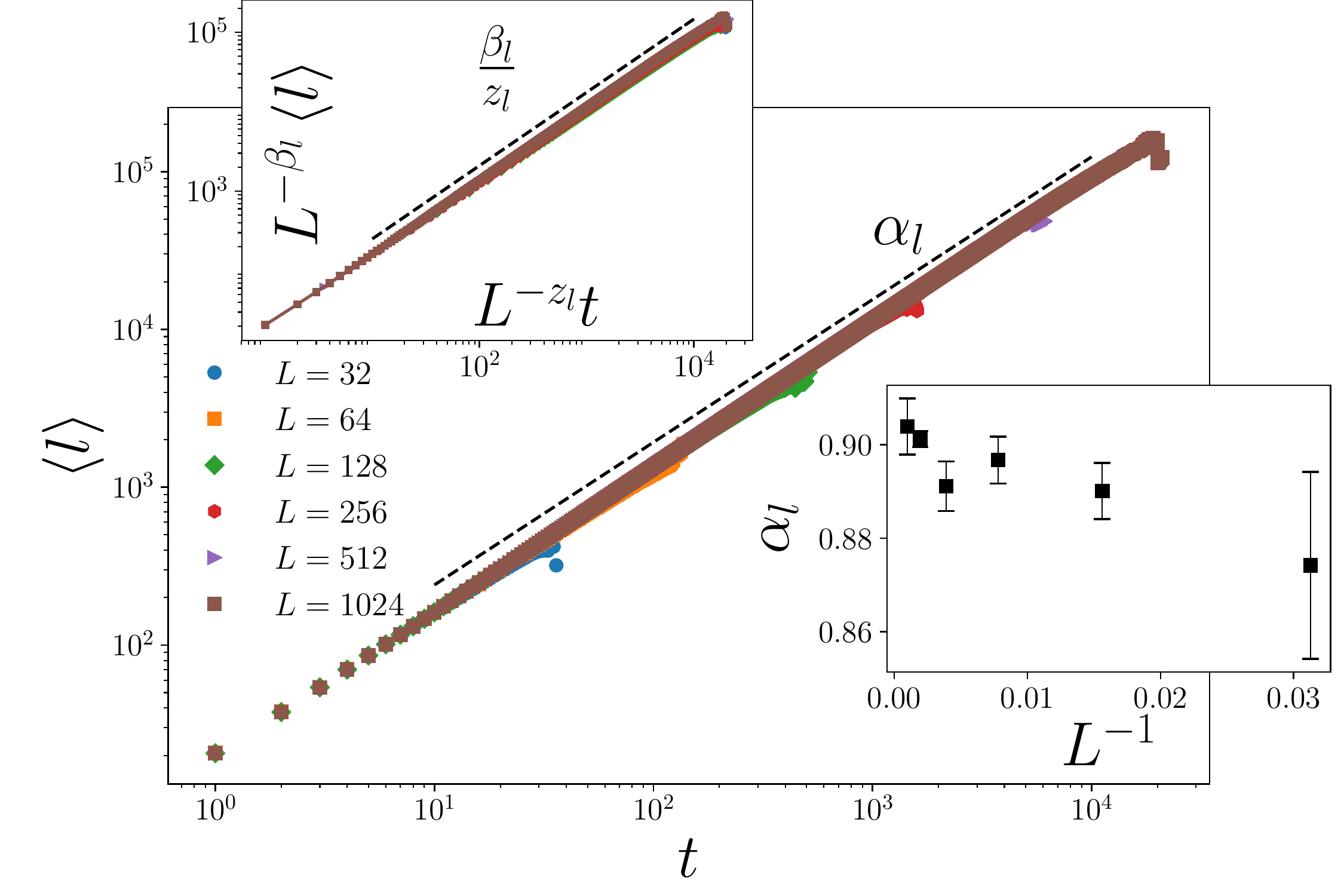}
		\caption{}
		\label{fig:t_l}
	\end{subfigure}
\caption{The time dependence of the average of (a) $r_m$, (b) $w$, (c) $r_l$, and (4) $l$ on time $t$. The upper insets are the data collapse analysis, showing the corresponding $\beta/z$ exponent. The lower insets are the $L$ dependence of the slopes of the curves at the main panels, to be extrapolated to $L\rightarrow\infty$. }
	\label{Fig:average}
\end{figure*}

\begin{figure}
	\centerline{\includegraphics[scale=.38]{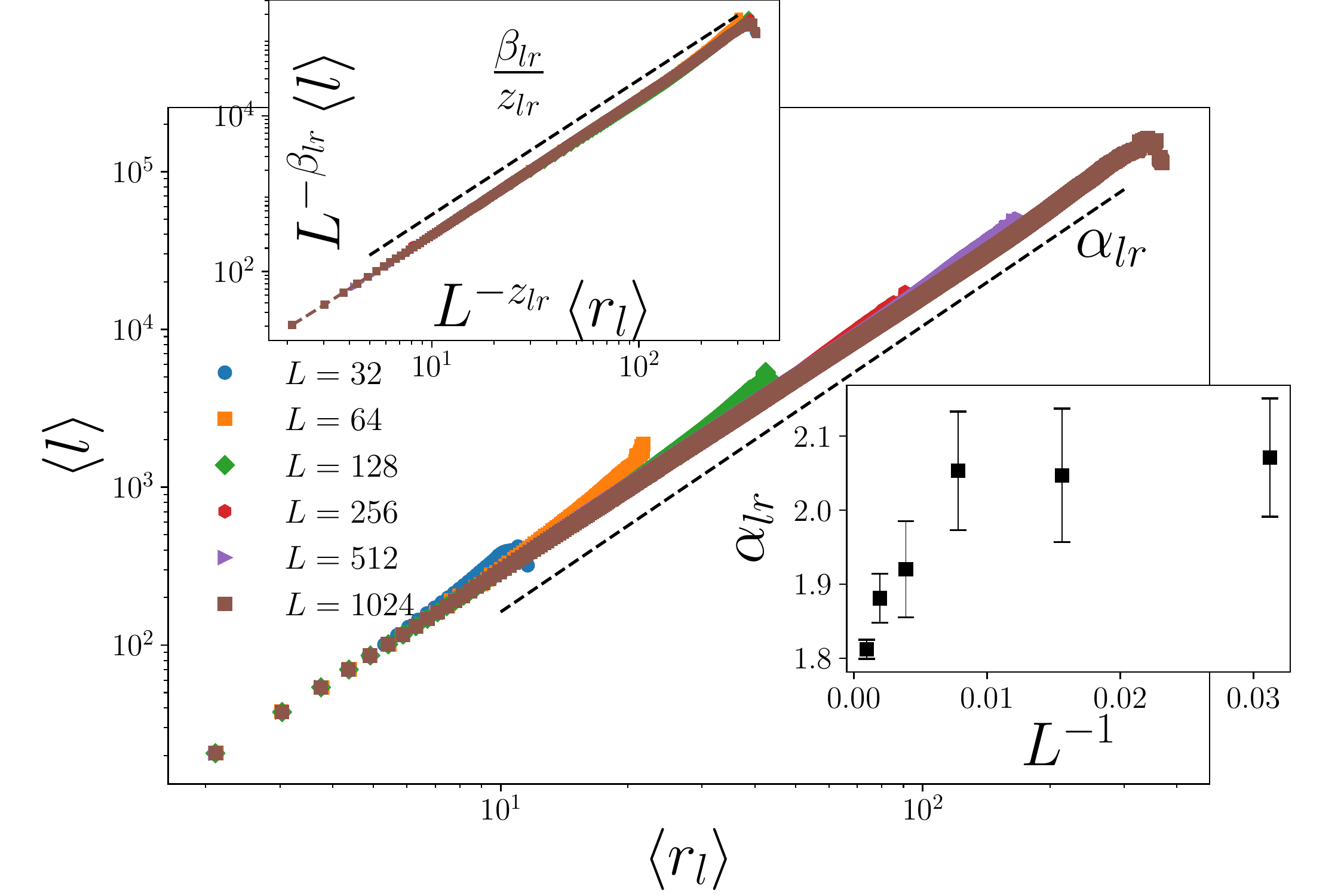}}
	\caption{Log-log plot of $l-r_l$ graph for \textit{all times}, whose slope is the dynamical fractal dimension. in the inset the fractal dimension is shown in terms of $1/L$, and in the upper inset the data collapse of the data is presented, from which the exponents $\beta_{lr}$ and $z_{lr}$ are extracted.}
	\label{fig:rl_l}
\end{figure}

\section{Results}\label{SEC:results}
To control the finite size effects, we simulated the model for square lattices with sizes $L=64,128,256,512,1024$, and measured the time dependence of $r_m$, $r_l$, $l$, and $w$. In the Fig.~\ref{Fig:average} we show the time dependence of $\left\langle x(t) \right\rangle $ in terms of $t$, $x=r_m,r_l,l$ and $w$. The graphs fit nicely by power-law for early times and change behavior at some crossover time to long time regime. This crossover takes place for all studied quantities. To quantify this behavior, we used data collapse technique based on finite-size scaling:
\begin{equation}
\begin{split}
\left\langle x(t,L)\right\rangle &=L^{\beta_{x}}F_x\left(\frac{t}{L^{z_x}} \right) \\
&=t^{\frac{\beta_x}{z_x}}G_x(\frac{t}{L^{z_x}})
\end{split}
\label{Eq:FSS}
\end{equation}
where $F_x$ and $G_x$ defined by $F_x(y)=y^{\frac{\beta_x}{z_x}}G_x(y)$ are universal functions with the asymptotic behaviors: $\lim_{y\rightarrow 0}G_x(y)=cnt$. For long time limit, let us suppose that $F_x(y)\propto y^{\gamma_x}$, so that $\left\langle x(t) \right\rangle\propto L^{\beta_x-z_x\gamma_x}t^{\gamma_x} $. Requiring that $\left\langle r(t) \right\rangle|_{\text{large times}}\propto L $, for $r_l$ and $r_m$, one obtains $\gamma_r=\frac{\beta_r-1}{z}$, so that $\left\langle r(t) \right\rangle|_{\text{large times}}\propto Lt^{\frac{\beta_r-1}{z_r}}$, implying that $\beta_r=1$ (since OSC stops at that time). The data collapse analysis presented in the insets of Fig.~\ref{Fig:average} support the finite size scaling hypothesis Eq.~\ref{Eq:FSS}. Note also that for small times (compared to $t_{\text{BH}}$ and $t_{\text{perc}}$) $l \sim t^{\beta_l}$, and $r_l \sim t^{\beta_{r_l}}$, so that $l\sim r^{D_f}$, where $D_f=\frac{\beta_l}{\beta_{r_l}}=1.8\pm 0.1$ is the dynamic fractal dimension. Another method to estimate the dynamic fractal dimension $D_f$ is within investigating directly $l$ and $r_l$ in any time, which is done in~Fig.~\ref{fig:rl_l} suggesting that $D_f\simeq 1.86\pm0.05$. The data collapse analysis of $l-r_l$ graphs is shown in the inset of this figure, confirming the above estimation of the fractal dimension $D_f=\frac{\beta_{lr}}{z_{lr}}=1.76\pm 0.05$. This fractal dimension should not be confused with the mass fractal dimension which is estimated to be $\simeq 1.82$ for low coordination numbers, and $\simeq 1.896$ for high coordination numbers~\cite{knackstedt2002nonuniversality}. The exponents are gathered in TABLE~\ref{tab:exponents}.  \\

\begin{table}
	\caption{The critical exponents $\beta$, $z$, $\frac{\beta}{z}$, and $\alpha$ of $r_l$, $r_m$, $l$ and $w$. The $(x,y)$ shows that data collapse analysis of the quantity $x$ in terms of $y$. The exponents $\beta$ and $z$ have been calculated using the data collapse method, and $\alpha$ was directly estimated by linear fitting of the quantities in terms of $t$ for all sizes, and extrapolating $L\rightarrow\infty$.}
	\begin{tabular}{c | c c c c c }
		\hline quantity & $(r_l,t)$ & $(r_m,t)$ & $(l,t)$ & $(w,t)$ & $(l,r_l)$ \\
		\hline $\beta$ & $0.97(9)$ & $1.00(2)$ & $1.65(8)$ & $1.00(2)$ & $1.70(5)$ \\
		\hline $z$ & $1.83(1)$ & $1.85(1)$ & $1.79(9)$ & $1.85(8)$ & $0.98(5)$\\
		\hline $\frac{\beta}{z}$ & $0.53(2)$ & $0.54(1)$ & $0.92(2)$ & $0.53(8)$ & $1.74(2)$ \\
		\hline $\alpha$ & $0.51(4)$ & $0.52(0)$ & $0.90(3)$ & $0.51(1)$ & $1.81(2)$ \\
		\hline
	\end{tabular}
	\label{tab:exponents}
\end{table}

\begin{figure*}
	\begin{subfigure}{0.45\textwidth}\includegraphics[width=\textwidth]{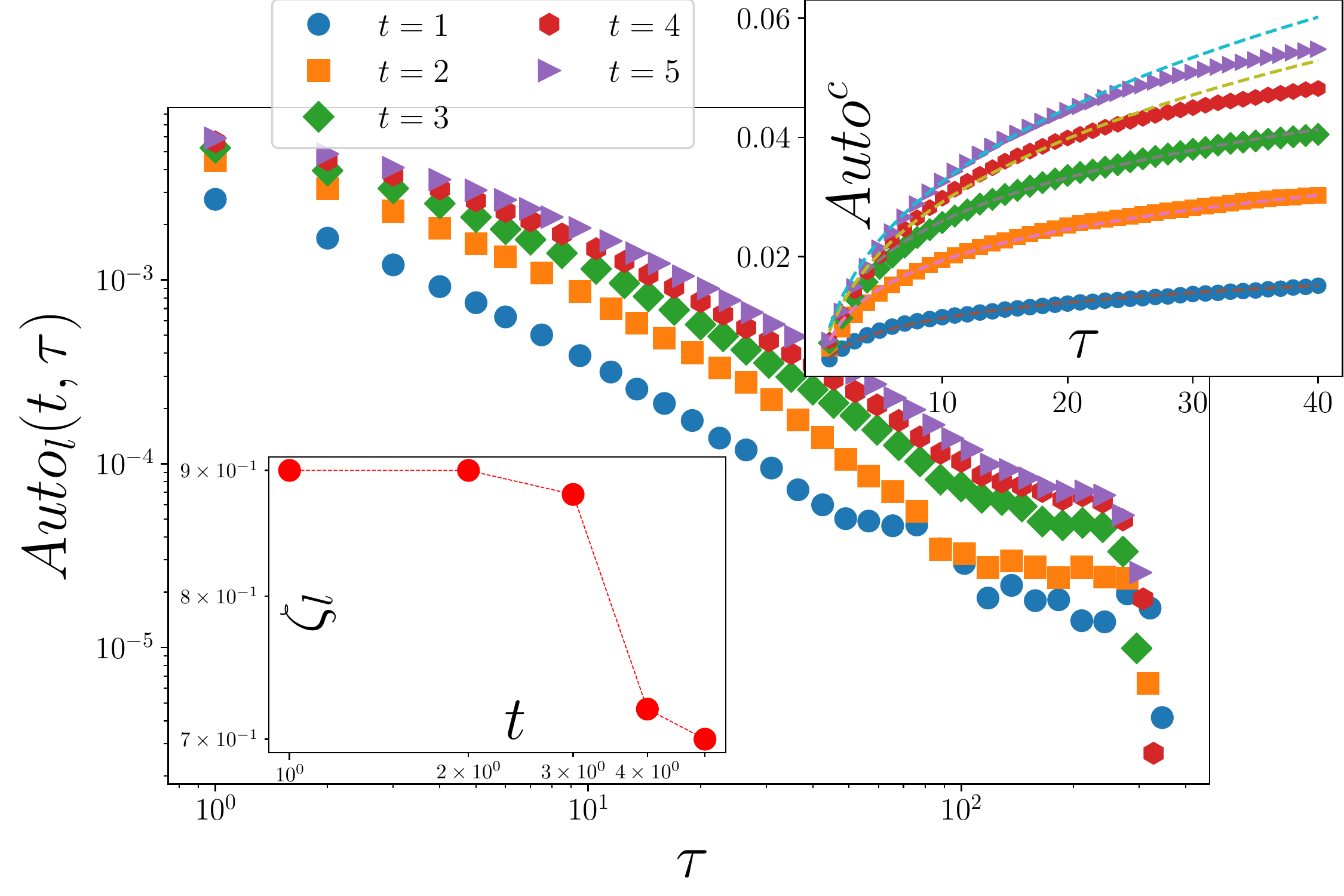}
		\caption{}
		\label{fig:harmonicflt}
	\end{subfigure}
	\centering
	\begin{subfigure}{0.45\textwidth}\includegraphics[width=\textwidth]{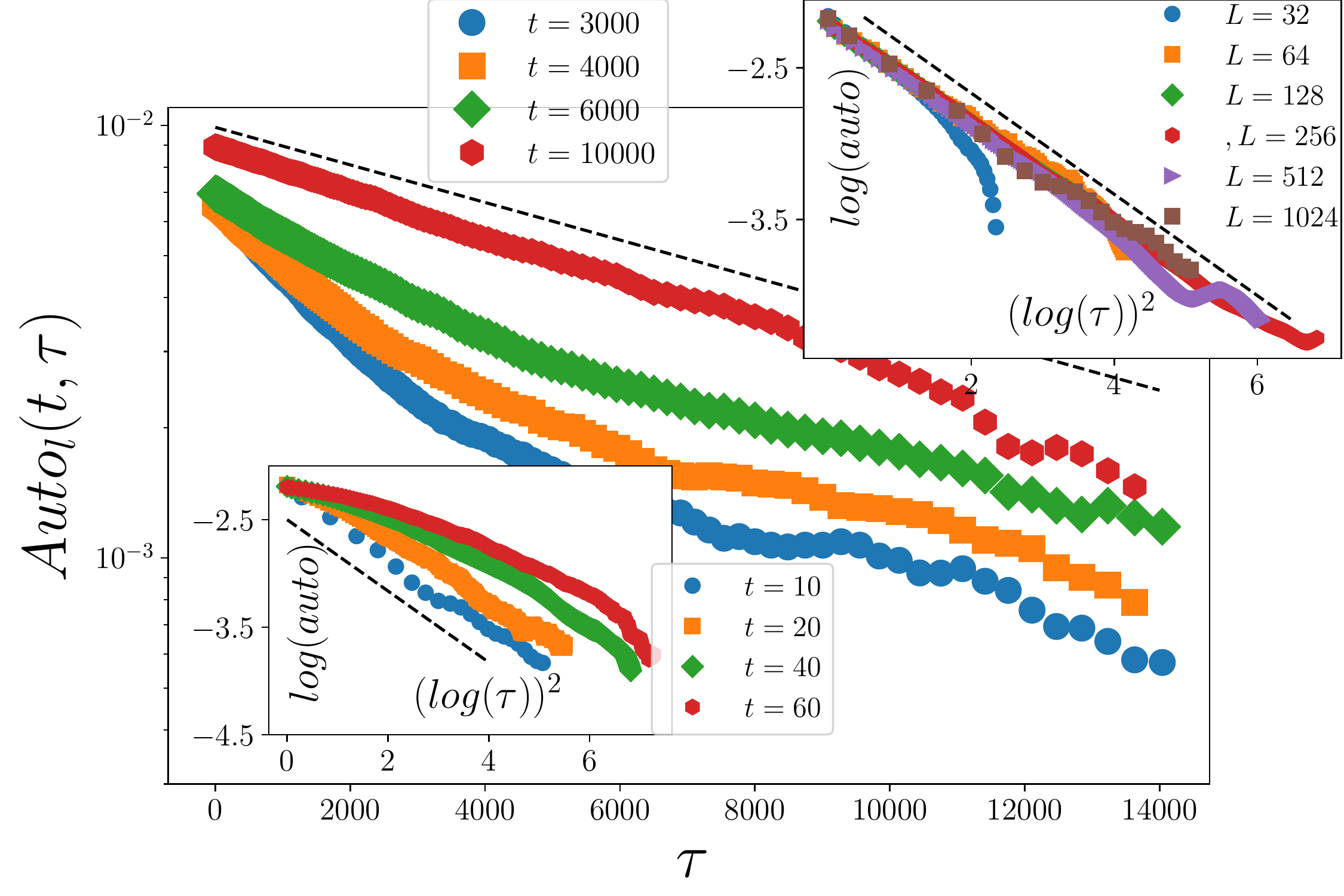}
		\caption{}
		\label{fig:large_timeflt1024}
	\end{subfigure}
	\centering
	\begin{subfigure}{0.45\textwidth}\includegraphics[width=\textwidth]{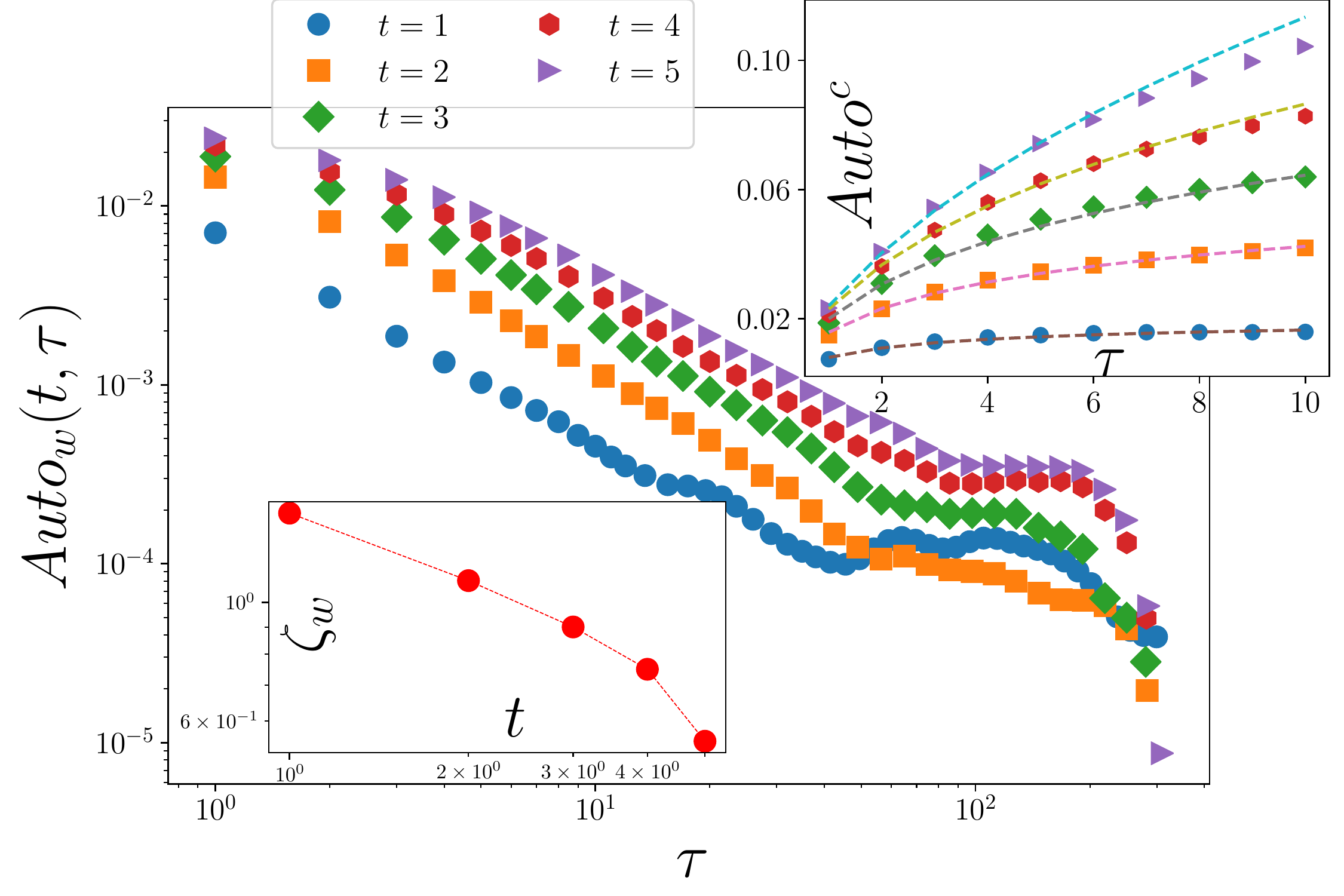}
		\caption{}
		\label{fig:hfw}
	\end{subfigure}
	\centering
	\begin{subfigure}{0.45\textwidth}\includegraphics[width=\textwidth]{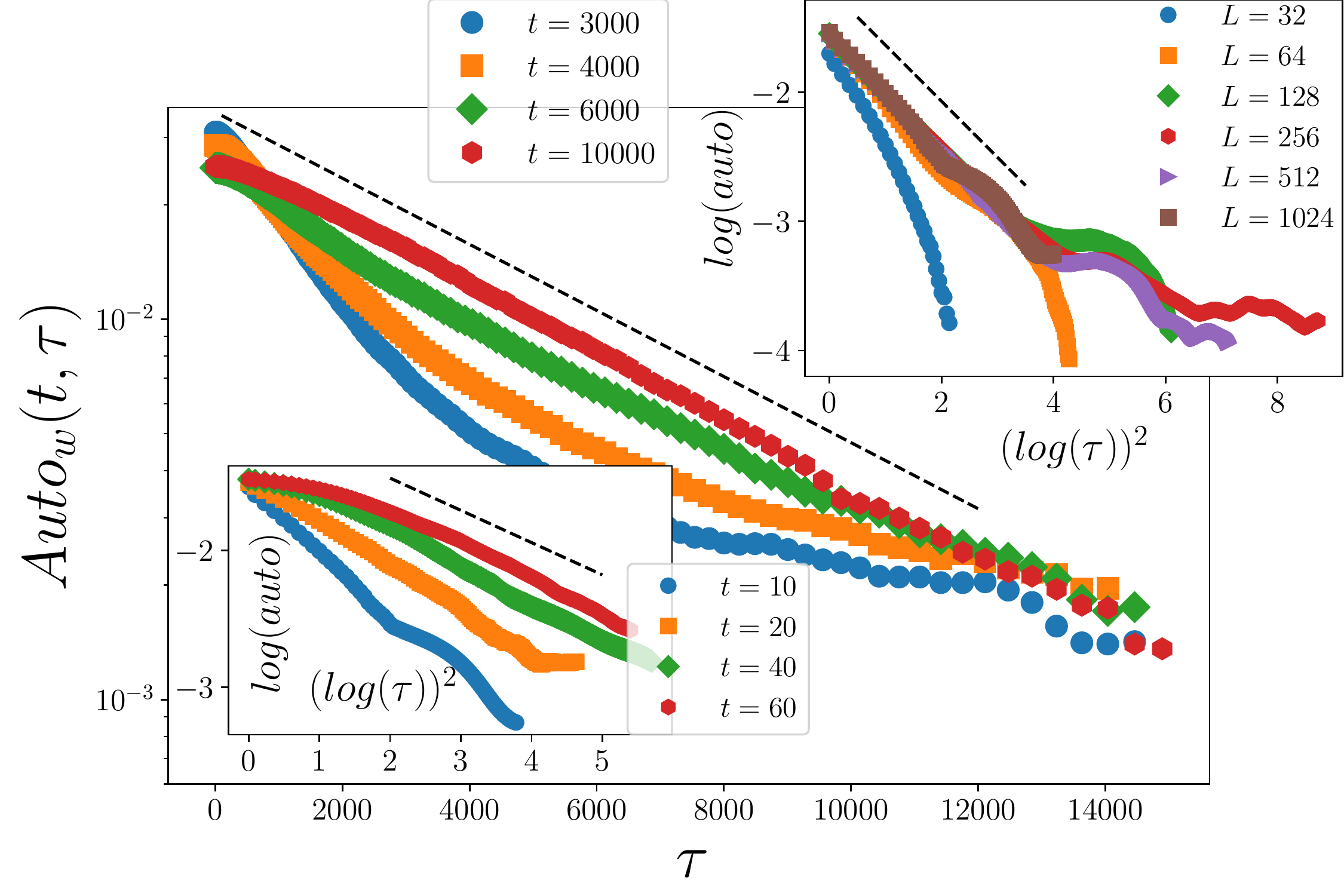}
		\caption{}
		\label{fig:large_timefwt1024}
	\end{subfigure}
	\centering
	\caption{$\tau$ dependence of $\text{Auto}_l$ ((a) and (b)), and $\text{Auto}_w$ ((c) and (d)). In (a) and (c) the plots are in log-log scale, and the lower insets show the accumulated autocorrelation functions, and in the upper insets, the exponents are reported in terms of $t$ in log-log scale. In (d) and (d) the main panels show the same in semi-log scale for very long times, and the insets show the mid-time behavior of $\log Auto$ in terms of $\left(\log \tau \right)^2$ for various times (lower insets) and system sizes (upper insets). The dashed lines are for eye guide for comparing the graphs with linear ones.}
	\label{Fig:auto}
\end{figure*}

To reveal the dynamical structure of the model, we consider the autocorrelation function of the quantities considered above, for which a crossover was observed. As stated in the previous section, $f(t,\tau)$ depends both on $t$ and $\tau$, representing the non-stationarity of the time series. We found that, in terms of $t$ there are three dynamical regimes that is depicted in Fig~\ref{Fig:auto}. These regimes are separated and identified by a crossover time $t_{\text{crossover}}$. For the early times $t\ll t_{\text{crossover}}$, the autocorrelations are power-law in terms of $\tau$ as shown in the main panel of Fig.~\ref{fig:harmonicflt}. For $t\approx t_{\text{crossover}}$ the autocorrelations are best fitted by log-normal functions, i.e. $\text{Auto}_x\propto \exp\left[-a_x\left(\log \tau \right)^2  \right] $ with non-universal exponent pre-factor $a_x$, and for large times $t\gg t_{\text{crossover}}$ the autocorrelation functions decay exponentially with $\tau$. The crossover time is estimated to be $t_{\text{crossover}}\approx 10$ for all lattice sizes, however its precise determination needs some additional works, and is beyond our analysis since we just looked at the autocorrelation functions. As is evident in the Fig.~\ref{fig:harmonicflt}, although the power-law fits are acceptable, they are noisy, and need another test that kills the noise. In the lower inset of this graph we show the accumulated autocorrelation function in terms of $\tau$ for various $t$ values, which contains considerable lower noise, and fit properly to Harmonic functions. The exponent of the power-laws $\zeta_l$ is shown in the upper inset which fixes at small times to $0.9\pm 0.01$, and decreases monotonically as time decreases. The fitting are valid up to $\approx t_{\text{crossover}}$ at which the power-law behavior is completely destroyed, and a new regime begins, that is shown in the lower and upper insets of Fig.~\ref{fig:large_timeflt1024}, where $\log \text{Auto}_l$ is shown as a function of $\left(\log \tau \right)^2 $. The finite size dependence at $t=t_{\text{crossover}}$ is shown in the upper inset, exhibiting a linear behavior as claimed. For long enough times (the main panel of Fig.~\ref{fig:large_timeflt1024}) the dependence is exponential. The same features were also observed for the roughness, as depicted in Figs.~\ref{fig:hfw} and~\ref{fig:large_timefwt1024}. We see that the quantities $f_x$, $x=r_l,l$, and $w$ uncover a dynamical crossover structure of IP, which should also be reflected in their increments, i.e. $\Delta f_x$ which in fact are the corresponding noises, whose correlations ($\text{Auto}_x^{\Delta}$) are important to realize the dynamical structure of the model. 

\begin{figure*}
	\begin{subfigure}{0.49\textwidth}\includegraphics[width=\textwidth]{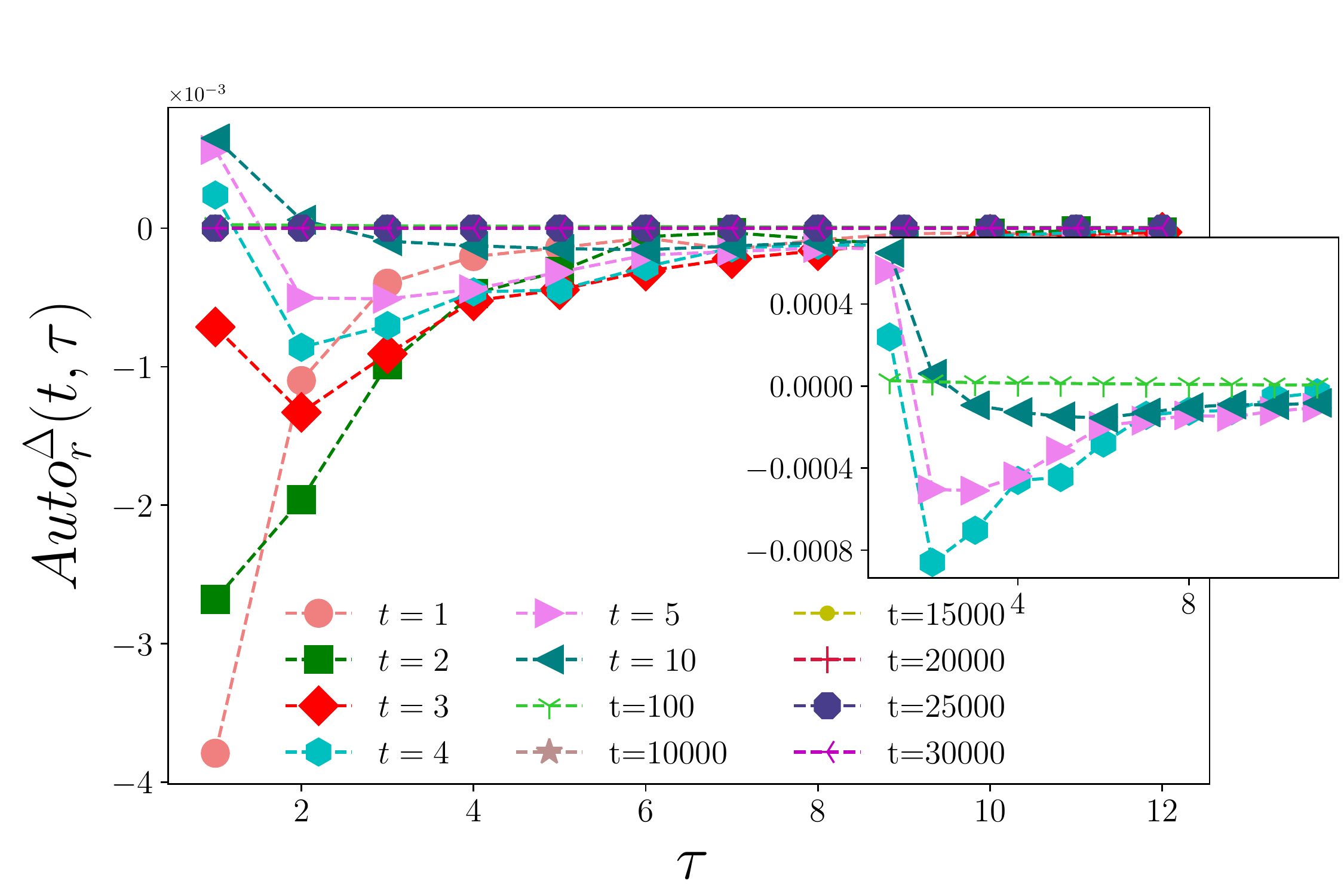}
		\caption{}
		\label{fig:zfr}
	\end{subfigure}
	\begin{subfigure}{0.49\textwidth}\includegraphics[width=\textwidth]{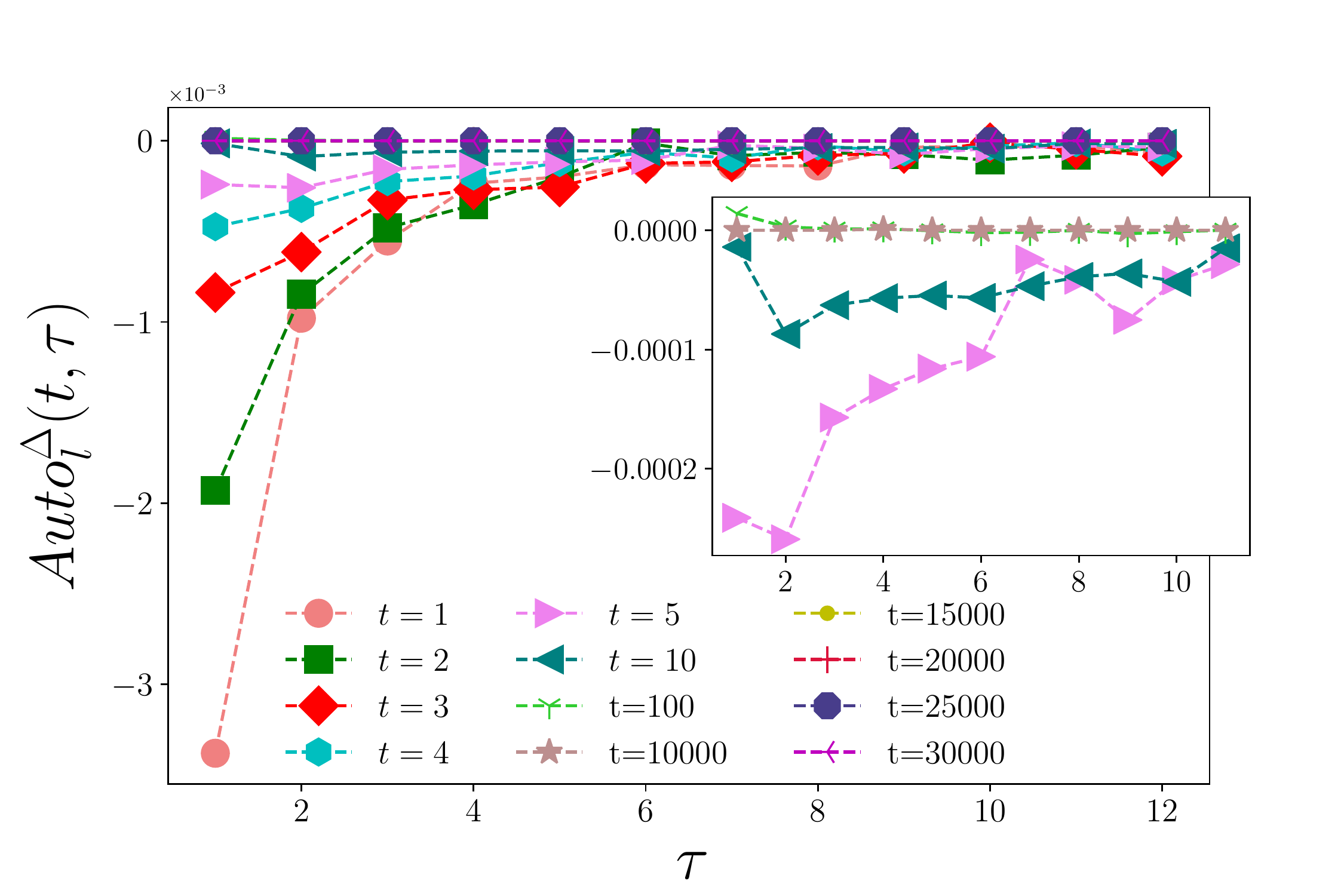}
		\caption{}
		\label{fig:zfl}
	\end{subfigure}
	\begin{subfigure}{0.49\textwidth}\includegraphics[width=\textwidth]{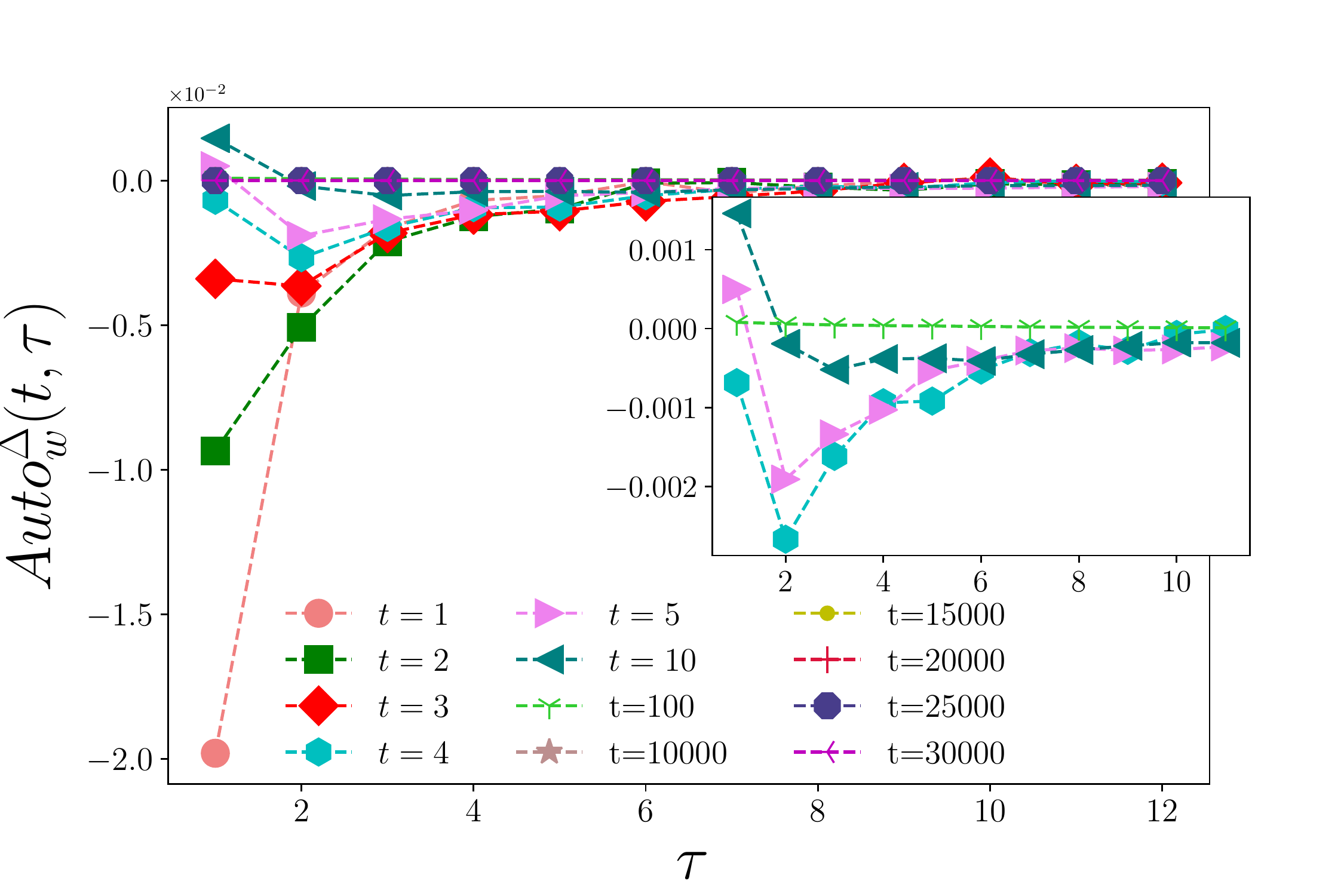}
		\caption{}
		\label{fig:zfw}
	\end{subfigure}
\caption{ The $\tau$ dependence of the autocorrelation functions of (a) $\Delta {f_r} $, (b) $ \Delta {f_l}$, and (c) $ \Delta {f_w}$ . (a)$ \Delta {f_l}$ for various rates of $t$ in the normal scale. The insets are magnified pictures of the main panels to show the anticorrelation/correlation crossovers.}
	\label{Fig:zfr_zfw_zfl}
\end{figure*}

In the Fig.~\ref{Fig:zfr_zfw_zfl} we show $\text{Auto}_x^{\Delta}(t,\tau)$ ($x=r_l,l,w$) in terms of $\tau$ for various amounts of $t$ for the maximum system size $L=1024$. For example let us focus on the $x=r_l$ (Fig.~\ref{fig:zfr}), for which an interesting change of sign takes place at small $\tau$s, signalling a anticorrelation-correlation (AC) crossover. This AC that occurs in the vicinity of $t=t_{\text{crossover}}$ has been observed for all system sizes that we considered in this work, and seems to survive at the thermodynamic limit, although some larger size simulations are necessary to make this hypothesis more precise and reliable. For $t\gtrsim t_{\text{crossover}}$ the correlations are positive. These behaviors are more evident in the zoomed graphs in the insets of these figures. The same features are also observed for $l$ (Fig.~\ref{fig:zfl}) and $w$ (Fig.~\ref{fig:zfw}). Another interesting behavior of these graphs is that for all graphs become negligibly small (more precisely become $\frac{1}{e}\times $ its maximum value) at $\tau_0\approx t_{\text{crossover}}$, showing that this is an intrinsic time scale in IP. Therefore $t_{\text{crossover}}$ is the average decay time (in terms of $\tau$) of the autocorrelations for all $t$s.

\section{Concluding Remarks}\label{SEC:concl}
In this paper we considered the two-dimensional invasion percolation (IP) with a focus on its dynamical properties. This problem is viewed as a growth process in which all statistical observables (the gyration radius $r_l$, $r_m$, the loop length $l$, and the roughness $w$) on the growing occupied sites clusters (OSC) are dynamic and time (defined as the integer part of $ \frac{m}{10} $) dependent. We calculated the fractal dimension of the external perimeter of the largest hole in the percolation time $t_{\text{percolation}}$, and confirmed that its fractal dimension is consistent with $D^I_f\approx \frac{4}{3}$ that corresponds to 2D self-avoiding walks (SAW) traces which is connected to percolation interfaces by the duality $\left( D^{H}_f-1\right) \left(D^I_f-1 \right)=\frac{1}{4}$~\cite{duplantier2000conformally}, ($D^{H}_f$ being the fractal dimension of the hulls). Using data collapse technique, we obtained the growth exponents of IP for all observables listed above. Importantly we analyzed the dynamic fractal dimension of the external perimeter of OSC (defined as the scaling exponent between $l$ and $r_l$ for all times) which, using the data collapse method, is estimated to be $1.76\pm 0.05$. The exponents are gathered in~\ref{tab:exponents}.\\

The dynamical features of IP was further investigated by analyzing the autocorrelation for $f_x\equiv x(t)/\left\langle x(t)\right\rangle -1$, which revealed that there is a crossover time $t_{\text{crossover}}$ which is nearly $L$-independent and is estimated to be $\approx 10$ (its more precise determination needs for other tests). Interestingly we observed that for $t\lesssim t_{\text{crossover}}$ the autocorrelations are power-law in terms of time difference $\tau=t'-t$, in the times in the vicinity of $t_{\text{crossover}}$ it behaves like log-normal functions, and for $t\gg t_{\text{crossover}}$ it decays exponentially. To be more precise, we analyzed also the autocorrelations between the increments $\Delta f_x(t)=f_x(t)-f(t-1)$, and observed that they also show a rich structure, i.e. the undergo a anticorrelation/correlation cross over at nearly the same times that the crossover between power-law/exponential decay occurs. The behaviors change not only in terms of the absolute time $t$, but also the time difference $\tau=t'-t$, e.g. $\tau_0=t_{\text{crossover}}$ is the average decay time of the autocorrelations.\\

This analysis suggests that there is an intrinsic crossover time scale $t_{\text{crossover}}$, at which IP changes its dynamical behaviors, importantly form power-law decay to exponential decay, and at the same time from anticorrelations (negative correlations) to positive correlations. Suppose that the behavior of $x(t)$ is controlled by a force, so that $\frac{1}{x}\frac{\delta x}{\delta t}= F_{\text{deter}}+\delta F_{\text{noise}}$, where $F_{\text{deter}}$ is the deterministic force, and $\delta F_{\text{noise}}$ is the noise part. Then 
\begin{equation}
\begin{split}
\frac{1}{\delta t}\delta \left( \frac{x}{\left\langle x\right\rangle }\right) &=\left\langle x\right\rangle^{-1}\frac{\delta x}{\delta t}-\frac{x}{\left\langle x\right\rangle^2}\frac{\delta \left\langle x\right\rangle }{\delta t}\\
&\simeq  F_{\text{deter}}+\delta F_{\text{noise}} -  F_{\text{deter}}=\delta F_{\text{noise}}\\
&
\end{split}
\end{equation}
where in the last line we used $\frac{x}{\left\langle x\right\rangle^2}\frac{\delta \left\langle x\right\rangle }{\delta t}\simeq F_{\text{deter}}$. We therefore see that $\Delta f_x(t)=\delta (\frac{x(t)}{\left\langle x(t)\right\rangle })=\delta F_{\text{noise}}$ (note that $\delta t\equiv 1$), i.e. $\Delta f_x(t)$ is nothing but the random force acting on $x$. At the crossover time, the autocorrelation of these random forces changes sign, i.e. for $t\lesssim t_{\text{crossover}}$, when the force becomes more than average, at the next time, the background effects compensate it, and vice versa. For $t\gg t_{\text{crossover}}$ however, the autocorrelation of this force becomes positive.
\bibliography{refs}

\end{document}